\documentclass[letterpaper, 10 pt, conference]{ieeeconf} 

\IEEEoverridecommandlockouts                              

\newcommand{\R}{{\mathbb R}}
\usepackage[mathscr]{euscript}
\usepackage{amsmath}
\usepackage{amssymb,amsfonts,enumerate}
\usepackage{graphics} 
\usepackage{epsfig} 
\usepackage{graphicx,color}
\usepackage{cite}
\usepackage{subfigure}
\usepackage{ushort}
\newtheorem {defi}{Definition}

\newtheorem {assum}{Assumption}
\usepackage[overload]{empheq}

\newcommand{\Z}{ \mathbb Z }

\newcommand{\rema}{\mathrm{rem}} 
\newcommand{\mymod}{\mathrm{mod}}
\newcommand{\myrem}{\mathrm{rem}}

\usepackage{pgf,tikz}
\usetikzlibrary{arrows}
\definecolor{aqaqaq}{rgb}{0.63,0.63,0.63}
\definecolor{qqzzcc}{rgb}{0,0.6,0.8}
\definecolor{dcrutc}{rgb}{0.86,0.08,0.24}
\definecolor{cqcqcq}{rgb}{0.75,0.75,0.75}
\definecolor{qqwuqq}{rgb}{0,0.39,0}
\definecolor{qqzzff}{rgb}{0,0.6,1}
\definecolor{xdxdff}{rgb}{0.49,0.49,1}
\definecolor{uuuuuu}{rgb}{0.27,0.27,0.27}
\definecolor{qqccqq}{rgb}{0,0.8,0}
\definecolor{ttffqq}{rgb}{0,.5,.5}
\definecolor{ffttww}{rgb}{0.8,0.8,0.8}
\definecolor{qqqqff}{rgb}{0,0,1}
\definecolor{rosso}{rgb}{1,0,0}

\begin{document}

\title{Formation control on Jordan curves based on noisy proximity measurements*\thanks{*Accepted for presentation at the 2019 European Control Conference}}
\author{Pietro DeLellis$^\dagger$\thanks{$^\dagger$Department of Electrical Engineering and Information Technology, University of Naples Federico II, 80125, Naples, Italy.},  Franco Garofalo$^\dagger$
 and Francesco Lo Iudice$^\dagger$
}

\maketitle

\begin{abstract}
The paradigmatic formation control problem of steering a multi-agent system towards a balanced circular formation has been the subject of extensive studies in the control engineering community. Indeed, this is due to the fact that it shares several features with relevant applications such as distributed environmental monitoring or fence-patrolling. However, these applications may also present some relevant differences from the ideal setting such as the curve on which the formation must be achieved not being a circle, or the measurements being neither ideal nor  as a continuous information flow. In this work, we attempt to fill this gap between theory and applications by considering the problem of steering a multi-agent system towards a balanced formation on a generic closed curve and under very restrictive assumptions on the information flow amongst the agents. We tackle this problem through an estimation and control strategy that borrows tools from interval analysis to guarantee the robustness that is required in the considered scenario.
\end{abstract}

\section{Introduction}

Formation control problems are becoming more and more relevant due to the increasing usage of unmanned vehicles in performing cooperative tasks across different domains \cite{BaAr:98,ohpa15}, surveillance, patrolling, and environmental monitoring problems being the main motivating applications \cite{pafr12,dhsu04,reca10}. In this context, the problem of ensuring a multi-agent system achieves a balanced circular formation has risen as the ideal testbed for different control strategies under diverse assumptions on the communication protocol among the agents\cite{ZhCh:13,SeWu:14,masi15,ChZh:11}. 
Decentralized approaches have been developed to achieve this control goal in the ideal case where each agent has full knowledge of its relative position with respect to its neighboring peers, and have proven flexible enough to cope with different static and time-varying communication topologies \cite{ChZh:11,KiSu:07,SmBr:05}. Later work have tackled the problem in presence of distance-only measurements \cite{jide17,DeGaLo:18}.

In its most general formulation, the problem of achieving a balanced circular formation consists of two tasks, that of approaching the circle and that of balancing the formation. Once the first task has been achieved, the agents can be seen as simple phase oscillators. One of the main limitations of this approach is that the curve along which the formation is achieved is restricted to be the circle. While this geometry is representative of the idea of controlling the formation of an ensemble of agents along a closed curve, indeed the main feature of surveillance and patrolling problems, it also introduces a critical simplification: Euclidean distance measurements correspond univocally to distances along the curve, thus facilitating the design of both the estimation and control strategies. Indeed, removing this simplification introduces uncertainty on the distance along the curve even in the ideal case in which measurements are not affected by noise, and are performed through sensors with an unlimited range.
In our recent work on this topic \cite{DeGar:13,DeGaLo:18} we have tackled the formation control problem from a different perspective, that is, relaxing the assumption that the information exchanged under the communication protocol can be uniquely associated to a relative position (or distance) along the curve on which the formation must be achieved. Namely, in \cite{DeGar:13} we have developed a distributed estimator that allows a phase oscillator to estimate its relative position with respect to its peers based on uncertain proximity measurements, that is, intermittent distance measurements obtained through a sensor with limited range. Then, in \cite{DeGaLo:18} we have leveraged this estimator to achieve a balanced circular formation when the range of the noisy sensors deployed is lower than the relative distance among the agents at steady-state, this implying a fully disconnected steady-state communication topology.

In \cite{DeGaLo:15}, we started to extend our approach to balance formation along generic Jordan curves under the simplifying assumption that i) the distance measurements were not corrupted by noise, and ii) the steady-state topology were connected. In this work, we remove these assumption and develop a formation control strategy that is (i) decentralized, (ii) flexible to cope with a restrictive communication protocol, and (iii) robust to measurement uncertainties introduced by non-ideal sensors and by the geometry of the problem. As a testbed, we consider the problem of achieving a balanced formation along a generic closed curve when the agents gather noisy measurements of their distance with respect to the agents that lie within the range of their sensors. We assume this range to be limited and such that, once a balanced formation is achieved, the communication topology becomes fully disconnected. Furthermore, we consider a futher element of uncertainty that typically arises in applications: if the distance between two agents is close to the sensor range, then a measure could or could not be available depending on the realization of a Bernoulli random variable. To tackle this problem, we design a symbiotic estimation and control strategy that allows each agent to recover, in distributed fashion, an interval estimate of its relative position along the curve with respect to its peers and then tune it to the desired value through a three level bang-bang controller. The effectiveness of our strategy is demonstrated numerically.

\section{Mathematical Preliminaries and notation}
First, we define $\mymod(z,m):= c$ as the remainder of $z$ modulo $m$, where $c$ is the unique solution of 
\begin{align*} 
c&=z-qm,\\ 0&\le c< |m|, q\in\Z.
\end{align*}
Also, we define $\rema(z,m) = \mymod(z-m/2,m)-m/2$.
Note that $-m/2 \le \mathrm{rem} (z,m) < m/2$ and 
$\rema( -z,m ) = -\rema(z,m)$, for all $z \neq k m$, $k\in\Z$.

Next, we introduce some operations and notation on intervals (\cite{Moore:09}):
\begin{itemize}
	\item given an interval $J \subset \mathbb{R}$, we denote its infimum $\ushort J$, its supremum $\bar J$, and its width $w(J) :=\bar{J}-\ushort{J}$;
	\item the Minkowski sum between two intervals $X,Y\in\R$ is $\lbrace x + y\ |\ x \in X, y \in Y \rbrace$. Note that the result is an interval $Z$ such that $\ushort Z = \min \lbrace \ushort X, \ushort Y\rbrace$ and $\bar Z = \min \lbrace \bar X, \bar Y\rbrace$.	
	\item given $\iota$ intervals $X_1,\ldots,X_\iota$, the infimum and the supremum of the interval hull $H=\mathrm{hull}_\lambda\left\{X_\lambda\right\}$ are given by $\ushort{H}=\underset{\lambda}{\mathrm{inf}} \lbrace \ushort{X}_{\lambda}\rbrace$ and 
	$\bar{H}=\underset{\lambda}{\mathrm{sup}} \lbrace \bar{X}_{\lambda}\rbrace$, respectively.
\end{itemize}
Finally, given a planar curve $\chi:[c_m ,c_M]\rightarrow\mathbb{R}^2$, we denote by $\chi_{[c_1 ,c_2]}$ its restriction to the interval $[c_1, c_2]$, with $c_1\ge c_m$ and $c_2\le c_M$. 
\section{Agent Dynamics and Control Goal}

We consider an ensemble of $N$ discrete-time integrators moving along a $C^0$ Jordan curve $\gamma:[0,l]\rightarrow \mathbb{R}^2$ parametrized by arclength, that is, for all $a\in[0,l]$, $\mathrm{arclength}\gamma_{[0,a]}=a$. Note that $\gamma$ maps positions along a one-dimensional curve in points on a two-dimensional Euclidean space. We express the dynamics of the generic $i$-th agent as 
\begin{equation}\label{eq:multi-agent}
p_i(k+1) = p_i(k)+s+u_i(k),
\end{equation}
where $p_i(k)-p_i(0)$ is the distance traveled along the curve $\gamma$ by the $i$-th agent up to time $k$. Hence, $s + u_i(k)$ is the distance traveled by agent $i$ in the sampling time, with $s$ being an intrinsic parameter of the multi-agent system, and $u_i(k)$ the deployed distributed control action. Without loss of generality, we assume that each agent's initial condition is encompassed in the interval $[0,l)$, and relabel the agents so that $p_i(0)>p_{i-1}(0)$, for all $i = 2,\dots,N$. To track the evolution of the difference between the distances traveled by the agents along the curve, we introduce the variables $p_{ij}(k) := p_i(k)-p_j(k)$, the dynamics of which are 
\begin{equation}\label{eq:rel_pos}
p_{ij}(k+1) = p_{ij}(k)+u_{ij}(k),  
\end{equation}
where $u_{ij}(k) := u_i(k)-u_j(k)$. As $p_{ij}(k)\in \mathbb{R}$, we define the agents' relative position on the curve $\gamma$ as
\begin{equation}\label{eq:x_def}
x_{ij}(k):=\myrem(p_{ij}(k),l)\in \left(-{l}/{2},{l}/{2}\right].
\end{equation}
While the agents considered in this work travel along a one-dimensional closed curve, we assume they are able to gather noisy measurements of their euclidean distance
\begin{align}\label{eq:alfa}
m_{ij}(k) :=\left\| \gamma(\mymod(p_j(k),l))-\gamma(\mymod(p_i(k),l))\right\|_2.
\end{align}
only if they lie sufficiently close to each other. Specifically, the measurement equation is given by
\begin{equation}\label{eq:out}
y_{ij}(k)=\beta\left(m_{ij}(k)\right)\left(m_{ij}(k)+v_{ij}(k)\right),
\end{equation}
where $v_{ij}(k)$ is the measurement noise, and $\beta(m_{ij}(k))$ is a Bernoulli random variable with parameter $q(m_{ij}(k))$ describing the probability of gathering a measurement at a given distance $m_{ij}(k)$. Namely,
\begin{equation}\label{eq:mes_a} 
q(m_{ij}(k))=
\begin{cases}
1, \qquad\quad\ \ \ \mbox{if}  \ m_{ij}(k) \in  [0,\ushort r],\\
0<\bar q<1, \  \mbox{if} \ m_{ij}(k) \in  \left(\ushort r, \bar r\right],\\
0, \qquad \quad \ \ \ \mbox{otherwise},\\
\end{cases}
\end{equation}
for all $i,j=1,\ldots,N$, $i\ne j$. Therefore, the absence of a measurement at time $k$ is represented by the output variable $y_{ij}(k)$ being 0. The output equation \eqref{eq:out} models the case in which a noisy proximity measurement of the Euclidean distance $m_{ij}(k)$ is surely available to two agents $i$ and $j$ only if they are sufficiently close, that is, if $m_{ij}(k)$ is encompassed in the closed interval $[0,\ushort r]$. Otherwise, we consider the realistic case in which, when $m_{ij}(k)$ is close to the nominal sensor range $r$, a measurement could or could not be available to the agents, respectively with probability $q\in [0,1]$ and $1-q$. Finally, we assume that when the agents are too far from each other, that is, $m_{ij}(k)>\bar r$, no measurement of their mutual distance is available.

Having introduced all the necessary notation, we are now ready to define the desired formation for our multi-agent system.
\begin{defi}\label{def:balanced}
We say that the multi-agent system \eqref{eq:multi-agent} achieves a $\varepsilon$-balanced formation along the closed curve $\gamma$ of length $l$ if, for all $i=2,\dots, N$,
\begin{equation}
\limsup_{k\rightarrow+\infty}|x_{ij}(k)-b|\leq \epsilon,
\end{equation}
for all $(i,j)\in\left\{(2,1),(3,2),\ldots,(1,N)\right\}$,
with $b:=l/N$.
\end{defi}
Note that this extends the standard definition in \cite{DeGaLo:18} to the case of a generic Jordan curve. 

Our goal is to design a decentralized control action $u_{ij}(k)$ that allows the multi-agent system \eqref{eq:multi-agent} to achieve a $\varepsilon$-balanced formation along a closed curve $\gamma$. Such a goal is made challenging by the fact that, to leverage a feedback control law, each agent must first estimate the position of its peers. The estimation is non-trivial as
\begin{itemize}
\item[(a)] the measurements are intermittent, and their collection is related to the distance among the agents in non-deterministic fashion, see equations \eqref{eq:out} and \eqref{eq:mes_a};
\item[(b)] when a measurement is gathered, it is affected by (bounded) noise;
\item[(c)] the relation between the measured quantity, the Euclidean distance $m_{ij}(k)$, and the agents' relative position on the curve is non injective. This is due both to the radius of curvature of $\gamma$ not being constant, and to the intrinsic ambiguity of distance measurements, which do not carry information on orientation. We emphasize that, even in the absence of measurement noise, any measurement $y_{ij}(k)$ would be compatible with a multi-interval set of relative positions on the curve $\gamma$.
\end{itemize}
To complete our problem statement, we now formally state the main assumptions underlying our work . 
\begin{assum}\label{assum:pos_ign}
Agents have no information on their absolute position on the curve $\gamma$.
\end{assum}
\begin{assum}\label{assum:noise_power}
The amplitude of the measurement noise in eq. \eqref{eq:mes_a} is bounded, that is, $|v_{ij}(k)|\leq \varphi.$
\end{assum}
\begin{assum}\label{assum:disc_form}
There do not exist two scalars $0\le s_1\le s_2$ such that $|s_1-s_2|\geq b$ and $||\gamma(s_1)-\gamma(s_2)||_2\leq \bar r$.
 \end{assum}
It is worth underlining that Assumption \ref{assum:disc_form} implies that the desired formation is fully disconnected, in the sense that at steady state no distance measurement is available to the agents.
\section{Estimation and Control Strategy}

We design a decentralized strategy to achieve a balanced formation on the curve $\gamma$ which prescribes that each agent $i$ combines the information provided by the output equation \eqref{eq:out} with knowledge of the dynamics and of the control law of its peers to (i) identify its closest follower $i-1$, and (ii) derive an estimate $\hat x_{i,i-1}(k)$ of the relative position $x_{i,i-1}(k)$, which is then fed back to a three level bang-bang feedback controller.

\begin{figure}\centering{
\includegraphics[trim= 0cm 0cm 0cm 0cm, scale=.85]{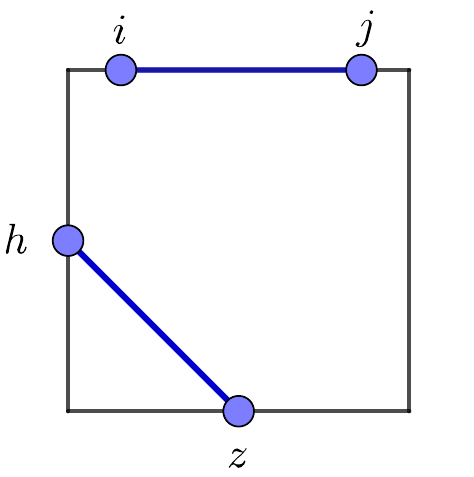}
\caption{An example of two pairs of agents $(i,j)$ and $(h,z)$ such that $m_{ij}=m_{hz}$ while $x_{ij}\neq x_{zh}$. The closed curve $\gamma$, the black solid line, is a square while the euclidean distances $m_{ij}$ and $m_{zh}$ are the blue solid segments.}\label{fig:sam_m_diff_x}}
\end{figure}

To explain how the agents exploit the information provided by each measurement $y_{ij}(k)$, let us point out that, taken altogether, Assumptions \ref{assum:pos_ign} and \ref{assum:noise_power} imply that each agent can associate to each value of $y_{ij}(k)$ a multi-interval set of values of $x_{ij}(k)$. This is due to several factors. Firstly, from eq. \eqref{eq:mes_a} and Assumption \ref{assum:noise_power}, whenever a measurement is available  (i.e. $\beta=1$) we can derive that
\begin{equation}\label{eq:inf_mes}
m_{ij}(k)\in [\max \lbrace y_{ij}(k)-\varphi,0\rbrace, \ \min\lbrace y_{ij}(k)+\varphi,  \bar r\rbrace].
\end{equation}
Secondly, as $\gamma(\cdot)$ maps points on a generic one-dimensional curve to points in $\mathbb{R}^2$, and as the euclidean distance does not carry information on orientation, there could be multiple values of $x_{ij}(k)$ compatible with the same value of $m_{ij}(k)$, as shown in the simple example of Fig. \ref{fig:sam_m_diff_x}. 

Indeed, from eq. \eqref{eq:mes_a}, we also know that if a measurement is not available (i.e. $\beta=0$), then 
\begin{equation}\label{eq:inf_no_mes}
m_{ij}(k)\in (\ushort r, \ l/2],
\end{equation}
which in turn translates into an interval knowledge on $x_{ij}(k)$. Summing up, regardless of whether a measurement is collected or not (regardless of the value of $\beta$), the knowledge of the measurement equations always allows agent $i$ to compute a multi-interval to which $x_{ij}(k)$ belongs. We denote by $\Upsilon^{ij}(k) = \lbrace \cup_l \Upsilon^{ij}_l(k)\rbrace$ such multi-interval. Combining $\Upsilon^{ij}(k)$ with knowledge of the dynamics and of an interval estimate of the control input of its peers, each agent $i$ produces a multi-interval interval estimate $\Gamma^{ij}(k|k) = \lbrace \cup_l \Gamma^{ij}_l(k|k)\rbrace $ of $x_{ij}(k)$. The equations governing the dynamics of this estimate are 
\begin{eqnarray}
\Gamma^{ij}(k|k-1) &=&\myrem (\Gamma^{ij}(k-1|k-1)+\hat u_{ij}(k-1),l,)\nonumber\\ \label{eq:est_b}
\Gamma^{ij}(k|k) &= &\Gamma^{ij}(k|k)\cap\Upsilon_{ij}(k),\\
\Gamma^{ij}(0|0) &=& \Upsilon^{ij}(0).\nonumber
\end{eqnarray}
Notice that, as agent $i$ cannot access the input of node $j$, the estimation strategy only relies on an estimate $\hat u_{ij}(k)$ of the relative input $u_{ij}(k)$. The estimation of $\hat u_{ij}(k)$ is facilitated by an appropriate choice of the control law $u_i$, $i=1,\ldots,n$. Specifically, we select a three-level control law that activates once an agent has identified univocally its closest follower. Therefore, before giving our control law, we have to give the following definition, where $H^{ij}(k|k)$ denotes the hull of the multi-interval $\Gamma^{ij}(k|k)$.
 \begin{defi}
Agent $i\ne 1$ identifies its closest follower at time $k_i$ if $k_i$ is the smallest integer ensuring there exists $k\leq k_i$ such that
\begin{subequations}\label{eq:follower}
\begin{equation}\label{eq:follower_a}
\ushort H^{i,i-1}(k|k)>0,
\end{equation}
\begin{equation}\label{eq:follower_b}
\bar H^{i,i-1}(k|k) < \ushort{\Gamma}_l^{ij}(k|k), \ \forall l |
\ushort{\Gamma}_l^{ij}(k|k)>0, \forall j \neq i-1.
\end{equation}
\end{subequations}
\end{defi}
Now, we are ready to give our control law. First, we randomly elect the pacemaker, w.l.o.g. agent $1$, whose control law is
\begin{equation}
u_1(k) = d \ \forall k.\label{eq:our_control_law_leader}
\end{equation}
The decentralized control action governing the motion of the other agents $i=2,\dots,N$ activates when they identify their closest follower, and then prescribes that $i$ is \emph{pushed} by its closest follower $i-1$, i.e.
\begin{subequations}\label{eq:contr_strat}
\begin{align}[left ={u_{i}(k) = \empheqlbrace}]
\label{eq:our_control_law_others_a}
d+ K\mathrm{sgn^+}( b- \hat x_{i,i-1}(k), \ &\forall k\geq k_i,\\
\label{eq:our_control_law_others_b}
0, \qquad \qquad \qquad \qquad  \qquad  \ &\mbox{otherwise},
\end{align}
\end{subequations}
where $x_{ij}(k)$ is the scalar estimate of $x_{ij}(k)$, computed as 
$\hat x_{ij}(k) = \ushort H^{ij}(k)$.
Taken altogether, equations \eqref{eq:our_control_law_leader} and \eqref{eq:contr_strat} imply that $u_i(k)\in \lbrace 0, \ d, \ d+K \rbrace$. Notice that, once agent $i$ identified its follower, it adjusts its speed to the maximal possible value $d+K$ until it reaches the desired spacing with agent $i-1$. Once  $b-\hat x_{i,i-1}(k)$ becomes negative, then agent $i$ becomes aware it has achieved its control goal, and thus starts traveling at the reference speed $d$. The general idea here, is that the pacemaker, which is the only agent such that $u_i(0)\neq 0$, ignites a mechanism in which agent $1$ pushes agent $2$ which in turn pushes agent $3$ and so on until each agent has achieve the desired spacing with its closest follower. This mechanism will eventually come to an end as the control law of the pacemaker lacks a feedback term.

This three-level control action allows a decentralized interval estimation of $\hat u_{ij}(k): =\hat u_j - u_i$, where $\hat u_j$ is the estimate made by $i$ of the speed of agent $j$. Specifically, before identifying its closest follower, 
\begin{equation}\label{eq:u_est_j}
\hat u_{j}(k) =  [0 \ d+K] \ \  \forall  k<k_i \ \mathrm{and} \ \forall j.
\end{equation}
Once agent $i$ has identified its closest follower, according to eq. \eqref{eq:our_control_law_others_a} it only needs an estimate $\hat x_{i,i-1}(k)$of $x_{i,i-1}(k)$, as at that point and it only needs to compute $\Gamma^{ij}(k|k-1)$ only for $j=i-1$. Hence, only $\hat u_{i-1}(k)$ is needed which agent $i$ obtains as
\begin{align}\label{eq:u_est_a}
\hat u_{i-1}(k) & =  [d \ d+K], \  \mathrm{if}  \ k_i\leq k<\bar k_i,\\\label{eq:u_est_b}
\hat u_{i-1}(k) & =  d, \qquad \qquad\mathrm{if}  \ k\geq\bar k_i,
\end{align}
where $\bar k_i$ is the time first instant greater than $k_i$ such that 
\begin{equation}\label{eq:barki_def}
H^{i,i-1}(k)\cap H^{i,i-1}(k_i) = \emptyset.
\end{equation}
The idea behind equations \eqref{eq:u_est_a} is that initially agent $i$ must perform estimates of all the other agents' control signal. This is true until $k=k_i$, when finally $i$ can focus only on its follower, $i-1$. For an agent to identify its closest follower, a relative motion must have taken place among the two. Hence, from equation \eqref{eq:our_control_law_others_a}, after $k_i$ agent $i$ can infer that $u_{i-1}(k)\geq d$. Finally, as $i$ is aware of the control law of its peers, it knows that, to distance its closest follower, $i-1$ must have already achieved the desired spacing with $i-2$ and thus, from eq. \eqref{eq:our_control_law_others_b} it must have switched to $u_{i-1} = d$. Hence, as soon as agent $i$ is aware it has started traveling faster than its follower $i-1$, which is the meaning of eq. \eqref{eq:barki_def}, it sets $\hat u_{i-1} = d$ as prescribed by eq. \eqref{eq:u_est_b}.

\section{Validation}
\subsection{Numerical setting}
\begin{figure}\centering{
\includegraphics[trim= 0cm 4cm 0cm 4cm, scale=.4]{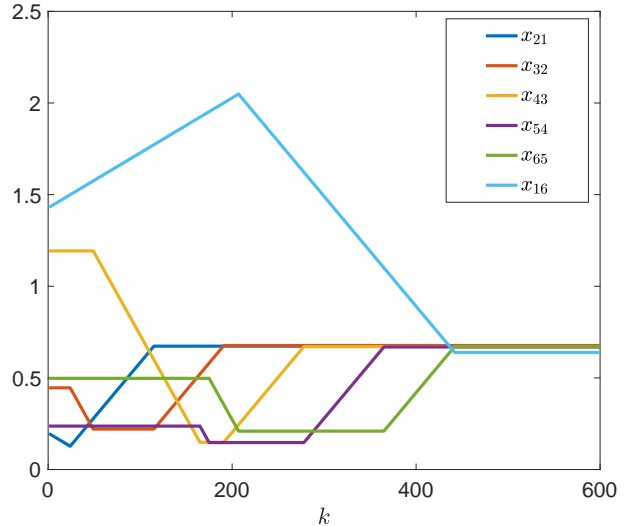}
\caption{Plot of $x_{ij}(k)$, $(i,j) = (2,1), \ (3,2), (4,3), \ (5,4), \ (6,5), (1,6$) for a representative simulation.}\label{fig:th_diff}}
\end{figure}

To validate our estimation and control strategy, we consider $N=6$ agents that must achieve an $\varepsilon$-balanced formation on a square of perimeter $l = 4$, which, from Definition \ref{def:balanced}, sets the value of the target spacing $b$ to $l/6$. Consistently with Assumption \ref{assum:disc_form}, we set $\ushort r = 0.32$ and $\bar r = 0.35$. We conclude the parametrization of equation \eqref{eq:mes_a} by selecting $\bar{q}=0.5$. Finally, the reference speed $d$ of the pacemaker is set to $0.003$. To facilitate the repeatability of our results, we now derive the expression of the multi-interval $\Upsilon^{ij}(k)$ as a function of $y_{ij}(k)$ for the specific shape of the selected curve $\gamma$. To do so, let us recall that, according to equation \eqref{eq:out} and from Assumption \ref{assum:noise_power}, if a measurement $y_{ij}(k)$ is available, then
$$m_{ij}(k)\in M^{ij}(k),$$
where
\begin{equation}
M^{ij}(k) :=[\max\lbrace 0, y_{ij}(k)-\varphi\rbrace,\min\lbrace y_{ij}(k)+\varphi, \bar r\}]\label{eq:mes2dist}
\end{equation}
Indeed, the relation between $M^{ij}(k)$ and $\Upsilon^{ij}(k)$ depends both on the specific curve $\gamma$ that is selected, and from $\bar r$. Having selected $\gamma$ to be a square with unit sides, and as $\bar r=0.35<1$, then $\Upsilon^{ij}(k)$ is made of two intervals. Moreover, under the considered parametrization it is trivial to show that $m_{ij}(k)\leq x_{ij}(k)\leq \sqrt{2}m_{ij}(k)$ and thus, if a measurement is available, we have
\begin{align}\label{eq:ups_for_square_mes}
&\Upsilon^{ij}(k)  = \Upsilon_1^{ij}(k) \cup \Upsilon_2^{ij}(k),\\
&\Upsilon_1^{ij}(k)= [-\sqrt{2}\bar{M}_{ij}(k), \ -\ushort{M}_{ij}(k)],\\
&\Upsilon_2^{ij}(k)= [\ushort{M}_{ij}(k), \ \sqrt{2}\bar{M}_{ij}(k) ].
\end{align} 
On the other hand, if a measurement is not available, then following the same lines of argument we can state that
\begin{align}\label{eq:ups_for_square_nomes}
&\Upsilon^{ij}(k)  = \Upsilon_1^{ij}(k) \cup \Upsilon_2^{ij}(k),\\
&\Upsilon_1^{ij}(k)= (\ushort{r}, \ l/2], \\
&\Upsilon_2^{ij}(k)= (-l/2, \ -\ushort{r}).
\end{align}
This estimation strategy, paired with the control strategy given in \eqref{eq:our_control_law_leader}-\eqref{eq:contr_strat} is effective in achieving an $\varepsilon$-balanced formation, with $\varepsilon$ being at most $0.09b$ across our numerical analyses. Figure \ref{fig:th_diff} shows the results of a simulation that is representative of the rationale behind our control strategy which prescribes that each agent $i$ except the pacemaker is pushed by its closest follower $i-1$. As the reader may notice, agent $i=2$ is the first agent such that $x_{i,i-1}(k)$ reaches its steady-state value. This as, before $k_2$, we have that $u_{i,i-1}(k) = 0$ for all $i=3,\dots,N$. In turn, $k_2$ is the first time instant such that $u_{32}(k)<0$, and thus a cascade is triggered that ends when $x_{65}(k)$ reaches its steady-state value, which is also the time instant in which $x_{16}(k)$ settles as agent $1$ is the pacemaker. 

\subsection{Performance analysis}

In what follows, we further delve on the performance of our strategy by assessing the role of the control gain $K$ and of the bound on the measurement noise $\varphi$ on the formation accuracy $\varepsilon$ and on the speed of convergence. Specifically, we vary $K$ in the set $\lbrace 0.5d, \ d, \ 1.5d, \ 2d \rbrace$ and, for each value of $K$, we vary $\varphi$ in the set $\lbrace 2K, \ 3K, \ 4K \rbrace$. For each combination of $K$ and $\varphi$, we have performed a set of $100$ numerical simulations each of $T=5000$ time instants. To avoid overtaking among the agents, we have selected the $100$ different initial conditions randomly, but ensuring that $|x_{ij}(0)|\geq 4\varphi_{\max}+2(d+K_{\max})$, where $K_{\max}$ and $\varphi_{\max}$ are the maximal values of $K$ and $\varphi$ across our numerical campaign.

First, we start by pointing out that in all experimental conditions, we have experienced convergence to an $\varepsilon$-balanced formation, as defined in Definition \ref{def:balanced}, with $\varepsilon = 0.09b$, while the average formation error across all performed experiments was $0.02b$. To evaluate the influence on the formation error $\varepsilon$ of varying the control gain $K$ and the measurement noise bound $\varphi$, we perform a two-way ANOVA. As summarized in table \ref{tab:vareps_anova}, we obtain negligible $p$-values for both parameters, indicating that both factors have a statistically significant effect on $\varepsilon$. Specifically, the formation accuracy improves when either $K$ or $\varphi$ is decreased.
\begin{table}
\vspace{2mm}
  \begin{center}
    \begin{tabular}{ccc}
    \hline
    Factor & degrees of freedom & $p$-value \\
    \hline
 	K    & 3  & 0  \\
 	$\varphi$    & 2  & 0\\
 	\hline         
    \end{tabular}%
    \vspace{2mm}
  \caption{Two-way ANOVA to test the influence on $\varepsilon$ of the factors $K$, and $\varphi$.}
  \label{tab:vareps_anova}
  \end{center}
\end{table}

The same statistical analysis was performed to test the influence of $K$ and $\varphi$ on the settling time $k_{5}$, which we define as the first time instant such that 
$$|x_{ij}(k)-x_{ij}(T)| \leq 0.05  x_{ij}(T),$$
for all $(i,j)\in\left\{(2,1),(3,2),\ldots,(1,N)\right\}$, $k\ge k_{5}$.

As summarized in Table \ref{tab:t_ass_anova}, while we found statistical evidence that varying the control gain $K$ it is possible to tune the speed at which the agents achieve the desired formation, we registered no significant effect of the bound on the measurement noise.
\begin{table}
\vspace{2mm}
  \begin{center}
    \begin{tabular}{ccc}
    \hline
    Factor & degrees of freedom & $p$-value \\
    \hline
 	K    & 3  & 0  \\
 	$\varphi$    & 2  & 0.74\\
 	\hline         
    \end{tabular}%
    \vspace{2mm}
  \caption{Two-way ANOVA to test the influence on the settling time $k_5$ of the factors $K$, and $\varphi$.}
  \label{tab:t_ass_anova}
  \end{center}
\end{table}

Taken altogether, our results indicate that as the measurement noise increases, the accuracy of the formation worsens, while the control gain $K$ can be tuned to achieve the optimal trade-off between the time in which the agents achieve the formation and its accuracy. We stress that, as we have tackled the problem in discrete time, the control gain $K$ can be reduced by reducing the sampling time, which would allow increasing the formation accuracy without compromising the time required to achieve the control goal. Indeed, this would come at the price of increasing the cost of the control architecture.

\section{Discussion}
In this manuscript, we tackled the problem of balancing a formation of autonomous agents along a generic Jordan curve. Leveraging our previous work, we have devised an estimation and control strategy whose effectiveness is not limited to circular formations. The combination of a prediction/correction estimation strategy with a three-level bang bang controller also allowed to deal with non-ideal proximity sensors characterized by i) a limited range, ii) bounded uncertainty, and iii) a non-deterministic behavior when the distance is close to their maximum range. We evaluated the effectiveness of the proposed strategy on a testbed curve, the square, which among all the regular polygons (with the exception of the triangle), is the worst approximation of the circle. Interestingly, we found that in all the simulations a balanced formation was achieved, with the only caveat that the agents should not be too close at time 0 to avoid possible overtaking. Furthermore, we observed that the control gain $K$ can be used to regulate the trade-off between the convergence speed and the accuracy of the formation. This encouraging numerical results constitute a further incentive to search for i) analytical conditions guaranteeing the effectiveness of the strategy, and ii) explicit estimates of convergence speed and formation accuracy. Our future work will be devoted to find useful analytical cues that can facilitate control design.

\bibliographystyle{IEEEtran}


\end{document}